\documentclass[12pt,preprint,numberedappendix]{emulateapj}

\newcommand\bibinc{n}		
\usepackage{subeqnarray}
\usepackage{amsmath}
\usepackage{hyperref}
\usepackage{ulem}
\usepackage{stmaryrd}
\hypersetup{colorlinks=true,linkcolor=black,citecolor=blue,urlcolor=black}

\usepackage{natbib}
\def\tt{\bf   }

\newcommand{\Eq}[1]{Equation\,(\ref{#1})}

\newcommand{\Sec}[1]{Section~\ref{#1}}

\newcommand{\Fig}[1]{Figure~\ref{#1}}

\newcommand {\exocam} {{\ttfamily ExoCAM}}

\setcitestyle{citesep={,}}
\begin{document}

\slugcomment{Accepted at ApJ}

\shorttitle{Baroclinic Criticality in Exoplanet Atmospheres}
\shortauthors{Komacek, Jansen, Wolf, \& Abbot}

\title{Scaling Relations for Terrestrial Exoplanet Atmospheres from Baroclinic Criticality}
\author{Thaddeus D. Komacek$^{1}$, Malte F. Jansen$^{1}$, Eric T. Wolf$^{2}$, and Dorian S. Abbot$^{1}$} \affil{$^{1}$Department of the Geophysical Sciences, The University of Chicago, Chicago, IL, 60637 \\ $^{2}$Laboratory for Atmospheric and Space Physics, Department of Atmospheric and Oceanic Sciences, University of Colorado, Boulder, CO, 80309 \\
\url{tkomacek@uchicago.edu}} 
\begin{abstract}
The macroturbulent atmospheric circulation of Earth-like planets mediates their equator-to-pole heat transport. For fast-rotating terrestrial planets, baroclinic instabilities in the mid-latitudes lead to turbulent eddies that act to transport heat poleward. In this work, we derive a scaling theory for the equator-to-pole temperature contrast and bulk lapse rate of terrestrial exoplanet atmospheres. This theory is built on the work of \cite{Jansen:2013}, and determines how unstable the atmosphere is to baroclinic instability (the baroclinic ``criticality'') through a balance between the baroclinic eddy heat flux and radiative heating/cooling. We compare our scaling theory to General Circulation Model (GCM) simulations and find that the theoretical predictions for equator-to-pole temperature contrast and bulk lapse rate broadly agree with GCM experiments with varying rotation rate and surface pressure throughout the baroclincally unstable regime. Our theoretical results show that baroclinic instabilities are a strong control of heat transport in the atmospheres of Earth-like exoplanets, and our scalings can be used to estimate the equator-to-pole temperature contrast and bulk lapse rate of terrestrial exoplanets. These scalings can be tested by spectroscopic retrievals and full-phase light curves of terrestrial exoplanets with future space telescopes.
\end{abstract}
\section{Introduction}
\label{sec:intro}
\indent We are approaching an era in which the atmospheres of terrestrial exoplanets will be detectable with both space-based \citep{Kreidberg:2016aa,Morley:2017aa} and ground-based \citep{Snellen:2015aa,Baker:2019aa,Molliere:2019aa} observatories. Observations with the future space telescopes LUVOIR/HabEx/OST will constrain the molecular abundances and atmospheric structure of terrestrial exoplanets orbiting Sun-like stars \citep{Feng:2018aa}, and their reflectance as a function of orbital phase will be inverted to derive maps of the planetary surface \citep{Cowan2009,Kawahara2010,Cowan2013,Lustig-Yaeger:2018aa}. These observations will determine if theories developed to understand the climate of Earth apply throughout the wide parameter space of exoplanets. \\
\indent The potential temperature, $\theta$, is a measure of entropy, and represents the temperature that a parcel would have if it were adiabatically displaced to a standard reference pressure. The slope of the potential temperature with height depends on whether the atmosphere is unstable or stable to dry convection. If the atmosphere is unstable to dry convection, the potential temperature decreases with height ($d\theta/dz <0$), if the atmosphere is stable the potential temperature increases with height ($d\theta/dz > 0$), and if the atmosphere is neutral the potential temperature is constant with height ($d\theta/dz = 0$). In an atmosphere unstable to dry convection, turbulent motions relax the troposphere toward constant potential temperature, causing the temperature profile to be that of a dry adiabat \citep{Pierrehumbert:2010}. Earth's atmosphere is stable to dry convection, but near neutrality to moist convection in the tropics. \\
\indent On Earth, the equator-to-pole temperature contrast is approximately the same as the potential temperature contrast between the surface and the tropopause. This means that isentropes (contours of constant entropy or potential temperature) in Earth's atmosphere that lie at the surface at the equator slope up to the tropopause at the pole. This slope is such that Earth's atmosphere is marginally unstable to certain types of baroclinic instability, which lead to the storms that represent weather in Earth's mid-latitudes \citep{Charney:1967aa,Eady:1949aa,Vallis:2006aa,Showman_2013_terrestrial_review}. The criticality of the atmosphere to baroclinic instability can be characterized by a baroclinic criticality parameter, $\xi$, which is $\approx 1$ for marginally critical conditions, $> 1$ for baroclinically unstable circulation, and $<1$ for stable flow.  Earth's atmosphere happens to have $\xi \approx 1$ \citep{Stone:1978aa}. \\
\indent Based on GCM simulations, \cite{Schneider:1961aa} and \cite{Schneider:2006aa} argued that Earth's atmosphere is forced to this marginally baroclincally unstable state. However, in a series of papers, \cite{Jansen:2012a,Jansen:2012,Jansen:2013} showed theoretically and numerically that varying atmospheric properties can cause the atmosphere to adjust to different baroclinic criticality parameters. Additionally, \citeauthor{Jansen:2013} built upon previous quasi-geostrophic theory \citep{Held:1996aa,Held:2007aa} to derive scaling relations for the baroclinic criticality parameter $\xi$. 
The baroclinic criticality parameter is directly related to isentropic slopes in the atmosphere, which in turn are controlled by the ratio of the equator-to-pole and surface-to-tropopause potential temperature contrasts. Using an additional constraint on the relationship between the horizontal and vertical heat transport, \citeauthor{Jansen:2013} further propose separate scalings for the equator-to-pole and surface-to-tropopause potential temperature contrast.   \\
\indent In this work, we use the theory of \citeauthor{Jansen:2013} to derive scalings for how the equator-to-pole temperature contrast and the bulk lapse rate of Earth-like exoplanets depend on planetary parameters. We define the bulk lapse rate as the minimum of the potential temperature contrast from the surface to the tropopause and the potential temperature contrast over one scale height. We compare our theory to results from a sophisticated exoplanet GCM to show the applicability of our scalings to terrestrial exoplanet atmospheres. Similar GCMs have been used previously to study how the circulation of terrestrial exoplanets orbiting Sun-like stars depend on a broad range of planetary parameters \citep{Yang:2014,Kaspi:2014,Popp:2016,chemke:2017,Way:2017aa,Wolf:2017,Jansen:2018aa,Kang:2019aa,Kang:2019ab}. We focus on planets the size of Earth to isolate the dependence of baroclinic criticality on planetary rotation rate and surface pressure. We find that the theory of \citeauthor{Jansen:2013} applies throughout the regime of baroclinically unstable and marginally critical atmospheres of Earth-sized exoplanets. \\
\indent This paper is organized as follows. In Section \ref{sec:theory}, we build upon the work of \citeauthor{Jansen:2013} to develop scalings for the baroclinic criticality parameter as a function of planetary parameters and use this to predict the scaling of the equator-to-pole temperature contrast and bulk lapse rate with planetary parameters. We outline our numerical setup in Section \ref{sec:methods}, and then compare our GCM results to our theoretical scalings in Section \ref{sec:comp}. Lastly, we discuss the application of our results to interpret future exoplanet observations, describe the limitations of our theory, and state conclusions in Section \ref{sec:disc}.
\section{Theoretical scalings}
\label{sec:theory}
\subsection{Criticality parameter}
The baroclinic criticality parameter is related to the ratio of the equator-to-pole and surface-to-tropopause potential temperature contrasts, $\xi \sim \Delta_h \bar{\theta}/\Delta_v\bar{\theta}$, where the overbars represent a zonal average and $\Delta$ represents a difference taken across the troposphere. Given that the slope of atmospheric isentropes is $s = -\left(\partial \bar{\theta}/\partial y\right)/\left(\partial \bar{\theta}/\partial z\right)$, where $y$ is latitude and $z$ is height, we can relate the criticality parameter to the bulk slope of atmospheric isentropes as \citep{Jansen:2013}
\begin{equation}
\label{eq:crit}
\xi = s\frac{a}{H}\mathrm{,}
\end{equation}
where $a$ is the planetary radius and $H$ is the height of the tropopause. The derivation in \cite{Jansen:2013} assumes a Boussinesq approximation, which is appropriate only if the tropopause height is small compared to the atmospheric scale height. In the opposite limit, the relevant height scale $H$ in \Eq{eq:crit} is the atmospheric scale height (e.g., \citealp{Chai:2014aa}). In this work, we take $H$ to be the minimum of the tropopause height and scale height.  \\
\indent We define the length of an isentrope to be the smaller of the length over which it rises from the surface to the tropopause or one scale height. As a result, we can write the slope of isentropes as the ratio of their height to their length, that is $s = H/l$. As in \cite{Jansen:2013}, we scale the length of isentropes with the distance that baroclinic eddies can diffusively mix the atmosphere over a radiative relaxation timescale $\tau_\mathrm{rad}$,
\begin{equation}
l_\mathrm{diff} \sim \sqrt{\tau_\mathrm{rad}D_\mathrm{eddy}} \mathrm{,}
\end{equation}
where $D_\mathrm{eddy}$ is a characteristic eddy diffusivity. Now, we rewrite the slope of isentropes as
\begin{equation}
s \sim \frac{H}{\sqrt{\tau_\mathrm{rad}D_\mathrm{eddy}}} \mathrm{,}
\end{equation}
and plugging this expression for $s$ into \Eq{eq:crit} we find that the criticality parameter scales as
\begin{equation}
\label{eq:criteddy}
\xi \sim \frac{a}{\sqrt{\tau_\mathrm{rad}D_\mathrm{eddy}}}. 
\end{equation}
\indent To relate the eddy diffusivity $D_\mathrm{eddy}$ to basic-state properties of the atmosphere, we assume that the eddy diffusivity scales as $D_\mathrm{eddy} \sim U L_\mathrm{Rh}$, where $U$ is a characteristic eddy velocity. $L_\mathrm{Rh} \approx \sqrt{U/\beta}$ is the Rhines scale, which is the scale at which the growth of macroturbulent eddies is arrested by differential rotation, where $\beta =  df/dy = 2\Omega \mathrm{cos}\left(\phi\right)/a$ is the change in the Coriolis parameter $f$ with latitudinal distance, where $\Omega$ is the rotation rate. We then use scalings from \cite{Held:1996aa} that relate the Rhines scale to the length scale at which gravity waves are affected by rotation (the Rossby deformation length $L_\mathrm{d}$) as $L_\mathrm{Rh} \sim \xi L_\mathrm{d}$. The Rossby deformation length $L_\mathrm{d} \approx \left(\partial_z \sqrt{b}\right) H/f$, where $b \approx g \alpha {\left(\theta-\theta_0\right)}$ is buoyancy, $g$ is gravity, $\alpha$ is the thermal expansion coefficient of air, and $\theta_0$ is a reference potential temperature. Solving for the eddy diffusivity, we recover the scaling of \cite{Held:1996aa}:
\begin{equation}
\label{eq:eddy}
D_\mathrm{eddy} \sim \beta \left(\xi L_\mathrm{d}\right)^3 \mathrm{.}
\end{equation}
\\
\indent Inserting \Eq{eq:eddy} into \Eq{eq:criteddy} and assuming that the radiative relaxation timescale scales as $\tau_\mathrm{rad} \propto p/(gT^3)$ \citep{showman_2002}, where $p$ is pressure and $T$ is temperature, we arrive at a scaling for the criticality parameter as a function of planetary parameters:
\begin{equation}
\label{eq:xiscaling}
\xi \sim \xi_\varoplus \left(\frac{\Omega}{\Omega_\varoplus}\right)^{2/5} \left(\frac{p}{p_\varoplus}\right)^{-1/5} \left(\frac{H}{H_\varoplus}\right)^{-3/5} \left(\frac{a}{a_\varoplus}\right)^{3/5} \left(\frac{g}{g_\varoplus}\right)^{-1/10} \mathrm{,}
\end{equation}
where $\xi_\varoplus \sim 1$ is the criticality parameter on Earth and all parameters are normalized to their values on Earth. Note that \Eq{eq:xiscaling} assumes that  changes in $\partial_z \theta$ are relatively small, which is consistent with the results discussed below. \Eq{eq:xiscaling} suggests that faster-rotating planets with less massive atmospheres will be more unstable to baroclinic instabilities. 
\subsection{Equator-to-pole temperature contrast and bulk lapse rate}
\indent \cite{Jansen:2013} showed that the baroclinic criticality parameter can be linked to the equator-to-pole temperature contrast and bulk lapse rate, the latter of which is related to the surface-to-tropopause temperature contrast. As in \cite{Jansen:2013}, for the purposes of this scaling derivation we will approximate the radiative heating/cooling as a Newtonian relaxation of potential temperature to an equilibrium value over the radiative timescale, with $d\theta/dt = -\left(\theta-\theta_\mathrm{eq}\right)/\tau_\mathrm{rad}$. Then, the vertical eddy heat flux needed to balance diabatic heating/cooling scales as
\begin{equation}
\label{eq:feddyv1}
F_\mathrm{eddy,v} \sim H \frac{\Delta_v \bar{\theta} - \Delta_v \bar{\theta}_\mathrm{eq}}{\tau_\mathrm{rad}} \mathrm{,}
\end{equation} 
where $\Delta_v$ denotes a vertical contrast, taken from the tropopause to the surface ($\Delta_v \theta = \theta_\mathrm{tropopause} - \theta_\mathrm{surface}$). \\
\indent Similarly, the horizontal eddy heat flux has to balance the diabatic heating and cooling at low and high latitudes. This allows us to relate the horizontal eddy heat flux $F_\mathrm{eddy,h}$ and the horizontal temperature contrast taken from equator to pole ($\Delta_h\theta = \theta_\mathrm{equator} - \theta_\mathrm{pole}$):
\begin{equation}
\label{eq:deltah}
F_\mathrm{eddy,h} \sim -a \frac{(\Delta_h \bar{\theta} - \Delta_h \bar{\theta}_\mathrm{eq})}{\tau_\mathrm{rad}} \mathrm{.}
\end{equation}
Following \cite{Jansen:2013}, we assume that the eddy heat flux is directed along isentropes. Additionally, we relate the ratio of the horizontal to vertical potential temperature contrasts (which scales directly with the criticality parameter) to the isentropic slope as $\Delta_h\bar{\theta}/\Delta_v\bar{\theta} \sim sa/H$. This allows us to relate the vertical eddy heat flux to the horizontal eddy heat flux as
\begin{equation}
\label{eq:deltav}
F_\mathrm{eddy,v} \sim s F_\mathrm{eddy,h} \sim \frac{H}{a} \frac{\Delta_h \bar{\theta}}{\Delta_v \bar{\theta}} F_\mathrm{eddy,h}   \mathrm{.}
\end{equation}  
\indent Combining Equations (\ref{eq:feddyv1}), (\ref{eq:deltah}) and (\ref{eq:deltav}), we relate the vertical and horizontal potential temperature contrasts as:
\begin{equation}
\label{eq:algebra}
\Delta_v\bar{\theta}\left(\Delta_v\bar{\theta}-\Delta_v\bar{\theta}_\mathrm{eq}\right) = -\Delta_h\bar{\theta}\left(\Delta_h\bar{\theta}-\Delta_h\bar{\theta}_\mathrm{eq}\right) \mathrm{.}
\end{equation}
Substituting $\xi \sim \Delta_h\bar{\theta}/\Delta_v\bar{\theta}$ into \Eq{eq:algebra} and assuming a stable background atmosphere with $\Delta_v \theta_\mathrm{eq} \approx 0$, we relate the horizontal potential temperature contrast individually to the criticality parameter
\begin{equation}
\label{eq:deltahixi}
\Delta_h\bar{\theta} \sim \frac{\Delta_h\theta_\mathrm{eq}}{1+\xi^{-2}} \mathrm{.}
\end{equation}
Dividing through by $\xi$, we relate the bulk lapse rate to the criticality parameter as
\begin{equation}
\label{eq:bulklapse}
\Delta_v\bar{\theta} \sim \frac{\Delta_h\theta_\mathrm{eq}}{\xi+\xi^{-1}} \mathrm{.}
\end{equation}
\Eq{eq:bulklapse} has a maximum in bulk lapse rate for $\xi = 1$, which is the regime that Earth itself is in.  \\
\indent Substituting our scaling for the baroclinic criticality parameter as a function of planetary parameters from \Eq{eq:xiscaling} into \Eq{eq:deltahixi}, we write a scaling for the equator-to-pole temperature contrast as a function of planetary parameters:
\begin{equation}
\begin{aligned}
\label{eq:deltahscaling}
\Delta_h\bar{\theta} \sim \left(\Delta_h\theta_\mathrm{eq}\right) \bigg[1+ & \xi_\varoplus^{-2} \left(\frac{\Omega}{\Omega_\varoplus}\right)^{-4/5} \left(\frac{p}{p_\varoplus}\right)^{2/5} \\ & \left(\frac{H}{H_\varoplus}\right)^{6/5} \left(\frac{a}{a_\varoplus}\right)^{-6/5} \left(\frac{g}{g_\varoplus}\right)^{1/5}\bigg]^{-1} \mathrm{,}
\end{aligned}
\end{equation}
where $\left(\Delta_h\theta_\mathrm{eq}\right)$ 
 is the equilibrium equator-to-pole potential temperature contrast. Similarly, we substitute our scaling for the criticality parameter from \Eq{eq:xiscaling} into \Eq{eq:bulklapse} to find a scaling for the bulk lapse rate as a function of planetary parameters:
\begin{equation}
\begin{aligned}
\label{eq:deltavscaling}
\Delta_v\theta \sim & \left(\Delta_h\theta_\mathrm{eq}\right) \bigg[\xi_\varoplus \left(\frac{\Omega}{\Omega_\varoplus}\right)^{2/5} \left(\frac{p}{p_\varoplus}\right)^{-1/5} \left(\frac{H}{H_\varoplus}\right)^{-3/5} \\ & \left(\frac{a}{a_\varoplus}\right)^{3/5} \left(\frac{g}{g_\varoplus}\right)^{-1/10} + \xi_\varoplus^{-1} \left(\frac{\Omega}{\Omega_\varoplus}\right)^{-2/5} \left(\frac{p}{p_\varoplus}\right)^{1/5} \\ & \left(\frac{H}{H_\varoplus}\right)^{3/5} \left(\frac{a}{a_\varoplus}\right)^{-3/5} \left(\frac{g}{g_\varoplus}\right)^{1/10} \bigg]^{-1} \ \mathrm{.}
\end{aligned}
\end{equation} \\
\indent We expect that the equator-to-pole temperature contrast increases for faster rotation rates and smaller surface pressures, until the limit where $\xi \gg 1$ when the temperature gradient approaches its radiative equilibrium value. We expect that the bulk lapse rate increases with faster rotation rates and smaller surface pressures for $\xi < 1$, increases with slower rotation rates and larger surface pressures when $\xi >1$, and has a maximum near $\xi = 1$. 
Next, we will introduce our numerical simulations, which will then be compared to our analytic scaling predictions for the criticality parameter, equator-to-pole temperature contrast, and bulk lapse rate in \Sec{sec:comp}.
\section{Numerical Methods}
\label{sec:methods}
\begin{figure*}
\centering
\includegraphics[width=1\textwidth]{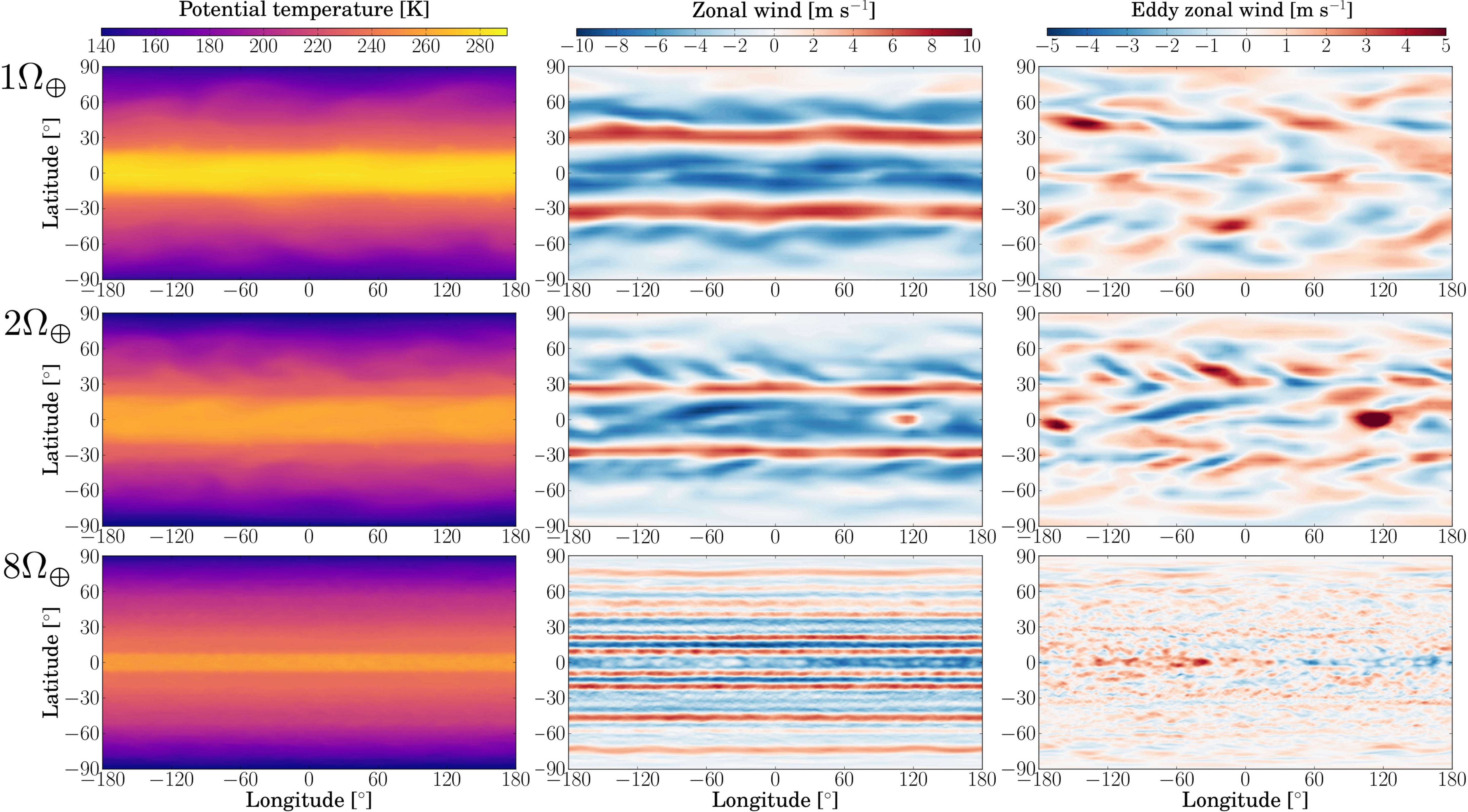}
\caption{ Potential temperature (left), zonal wind (center), and eddy component of the zonal wind (right) maps at the lowest atmospheric model level for rotation rates of $1, 2,~\mathrm{and}~8~\Omega_\varoplus$. The surface pressure for these simulations is 1 bar. Eddies that form from baroclinic instabilities are apparent in the mid-latitudes.}
\label{fig:T}
\end{figure*}
\indent To compare with our scaling theory, we perform simulations with the \exocam~GCM. \exocam~is a version of the Community Atmosphere Model version 4 with upgraded radiative transfer and water vapor absorption coefficients for application to exoplanets, and has been used for a wide range of exoplanet studies \citep{Kopp:2016,kopparapu2017,Wolf:2017,Wolf:2017aa,Haqq2018,Komacek:2019aa,Yang:2019aa}. In this work, we use the same basic model setup as \cite{kopparapu2017,Haqq2018,Komacek:2019aa}, and \cite{Yang:2019aa}: we consider aquaplanets without continents that have immobile slab oceans with a depth of $50~\mathrm{m}$ and an atmosphere comprised only of N$_2$ and H$_2$O. We consider planets with zero obliquity orbiting a Sun-like star, with varying rotation rates from $0.0625-8 \Omega_\varoplus$, where $\Omega_\varoplus$ is the rotation rate of Earth, and varying surface pressure from $0.25-4~\mathrm{bars}$. When varying rotation rate, we keep the surface pressure fixed at 1 bar. Similarly, when we vary surface pressure, we keep the rotation rate fixed to $1 \Omega_\varoplus$. \\
\indent These simulations are the same as those in \cite{Komacek:2019aa}, except that we extend the suite of simulations to faster rotation rates. All simulations assume an incident stellar flux of $1360.8 \ \mathrm{W} \ \mathrm{m}^{-2}$, equal to that of Earth. The majority of these simulations use a horizontal resolution of $4^\circ \times 5^\circ$ with 40 vertical levels and a timestep of $30~\mathrm{minutes}$. However, we use a horizontal resolution of $0.47^\circ \times 0.63^\circ$ and a timestep of $7.5~\mathrm{minutes}$ for the fastest rotating case, which has a rotation rate 8 times greater than that of Earth. \\
\indent \Fig{fig:T} shows maps of near-surface potential temperature, zonal wind, and eddy component of the zonal wind from a subset of our GCM simulations with varying rotation rates of $1-8 \Omega_\varoplus$. The eddy component of the zonal wind is $u' = u - \bar{u}$, where $u$ is the zonal wind and $\bar{u}$ is the zonal-mean of the zonal wind. We find that the width of tropical regions decreases and the number of zonal jets increases with increasing rotation rate, as expected from previous work \citep{Rhines:1975aa,Williams:1982aa,Kaspi:2014,Wang:2018}. We also find that the scale of mid-latitude eddies decreases with increasing rotation rate due to the decreasing deformation radius, as expected from previous work \citep{Schneider:2006aa,Kaspi:2011aa,Kaspi:2013aa,Wang:2018}. This basic understanding qualitatively agrees with that from our scaling theory, which found that the baroclinic criticality parameter should increase with increasing rotation rate.
\section{GCM-Theory comparison}
\label{sec:comp}
\subsection{Criticality parameter}
\begin{figure*}
\centering
\includegraphics[width=1\textwidth]{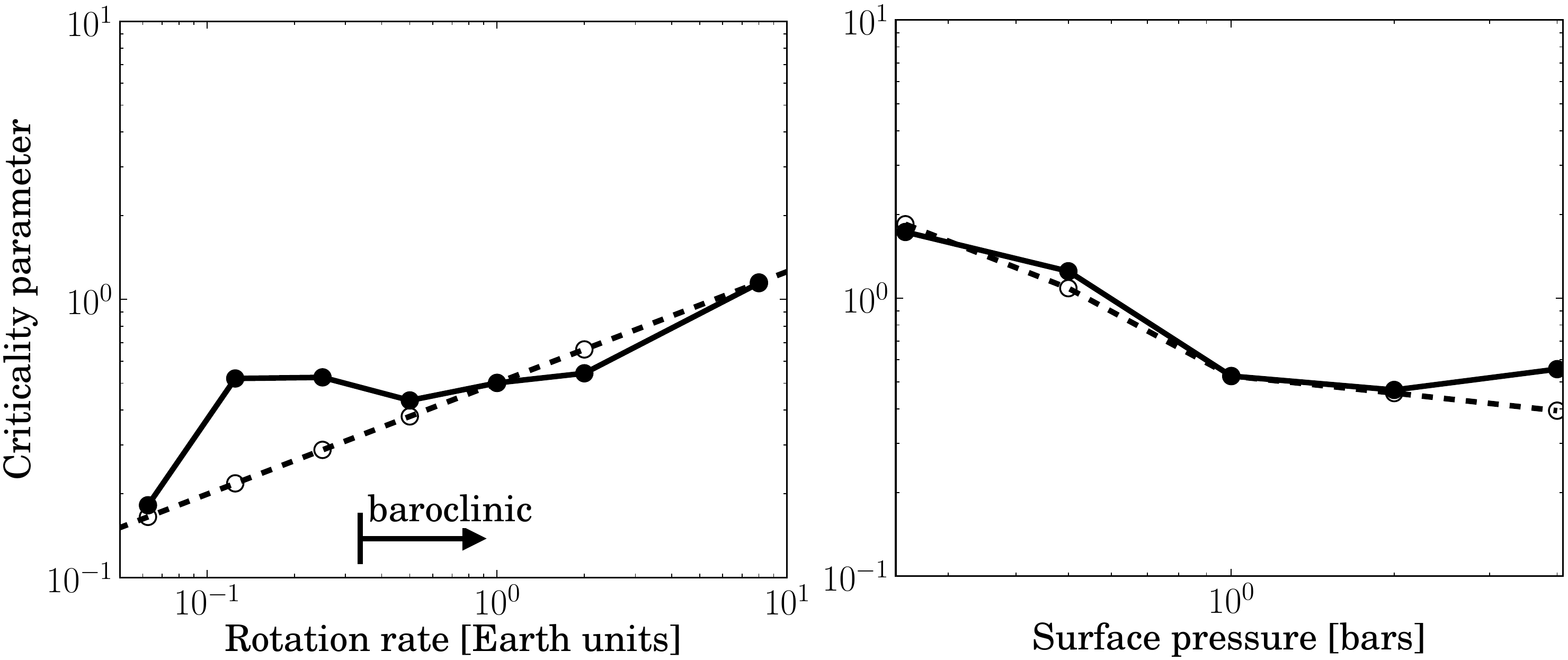}
\caption{Scaling for the criticality parameter (dashed lines) compared to GCM results (solid lines) for separately varying rotation rate (left panel) and surface pressure (right panel). For the simulations varying rotation rate (left panel), we keep the surface pressure fixed to 1 bar. For the simulations varying surface pressure (right panel), we keep the rotation rate fixed to that of Earth. We find good agreement between our scaling and GCM results throughout the baroclinically unstable regime, identified by the arrow. However, at rotation rates less than about $1/3 \Omega_\varoplus$ our scaling under-predicts the criticality parameter, as baroclinic effects weaken.}
\label{fig:crit}
\end{figure*}
\begin{figure*}[ht!]
\centering
\includegraphics[width=1\textwidth]{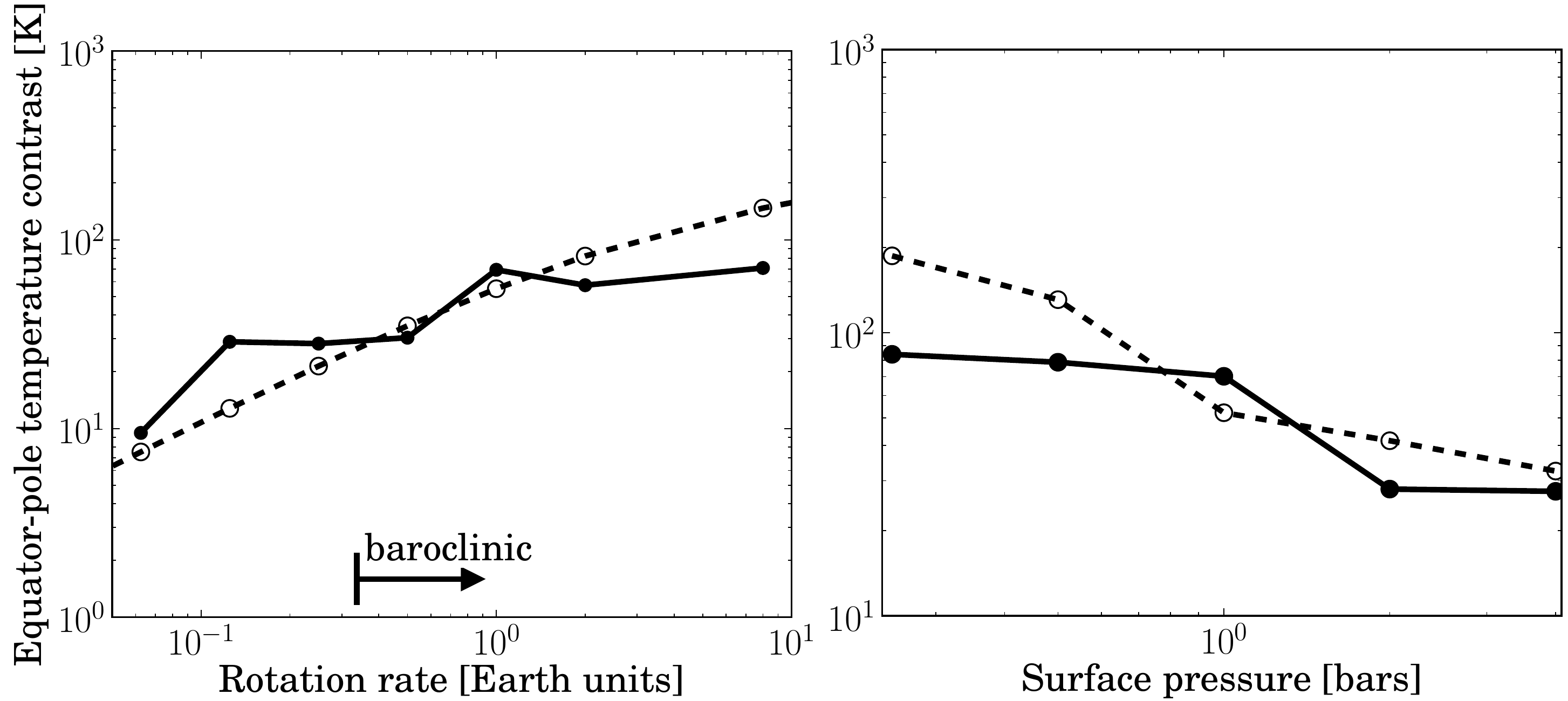}
\caption{Scaling for the equator-to-pole temperature contrast (dashed lines) compared to GCM results (solid lines) with varying rotation rate (left panel) and surface pressure (right panel). The equator-to-pole temperature contrast increases with increasing rotation rate and decreasing surface pressure in both our theory and GCM experiments. As a result, we find that our scaling explains the qualitative trends in equator-to-pole temperature contrast throughout the baroclinically unstable regime.}
\label{fig:delta}
\end{figure*}
\begin{figure*}[ht!]
\centering
\includegraphics[width=1\textwidth]{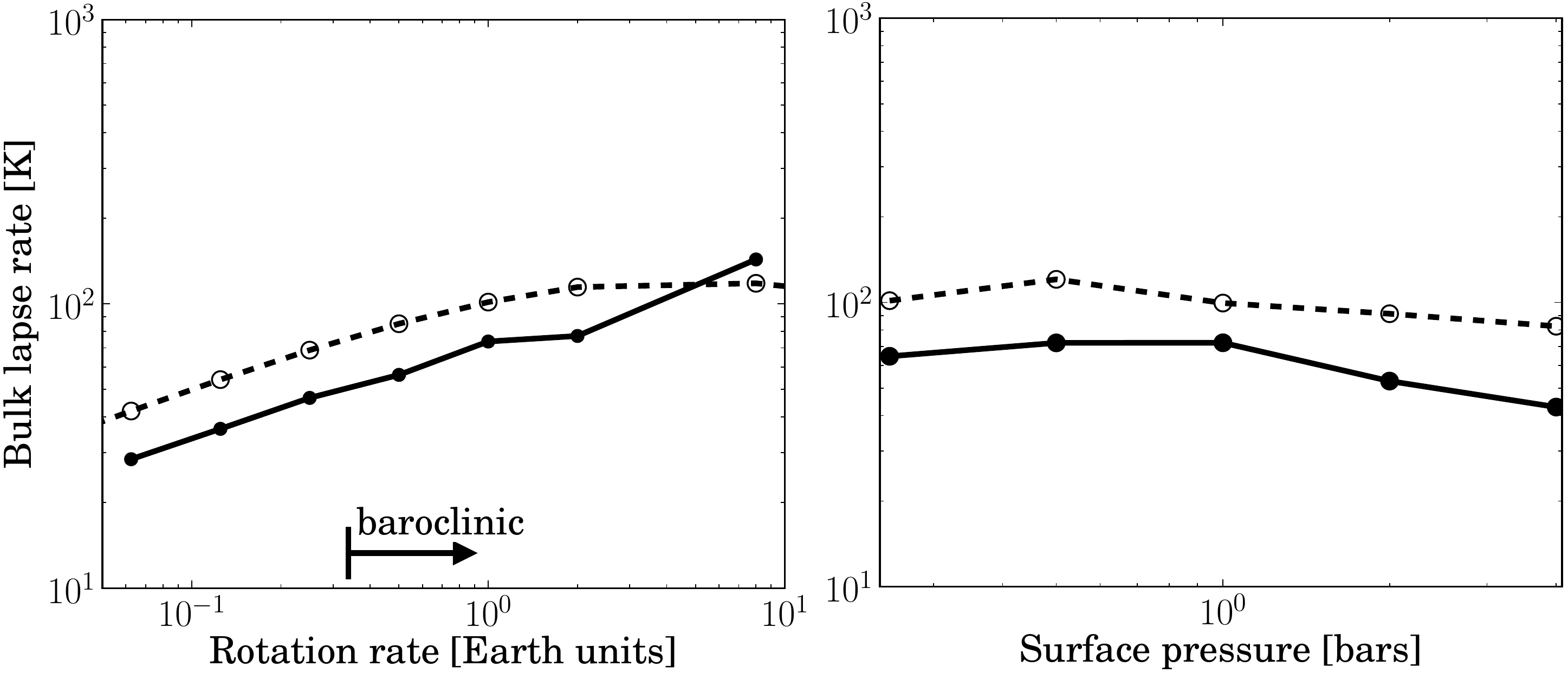}
\caption{Scaling for the bulk lapse rate (dashed lines) compared to GCM results for the mid-latitude bulk lapse rate with varying rotation rate (left panel) and surface pressure (right panel). Overall, the variations in bulk lapse rate with planetary parameters are relatively weak compared to the variations in equator-to-pole temperature contrast. As expected from our theory, the bulk lapse rate in our simulations increases with increasing rotation rate up to a maximum at $\approx 8 \Omega_\varoplus$. The bulk lapse rate increases and then decreases with increasing surface pressure in our simulations, as predicted by our theory.}
\label{fig:bulk}
\end{figure*}
\indent We calculate the baroclinic criticality parameter from our GCM results similarly to \cite{Jansen:2013}:
\begin{equation}
\label{eq:criticalityfromgcm}
\xi = \frac{a\left<\partial \bar{\theta}/\partial y\right>}{H\left<\partial \bar{\theta}/\partial z\right>} \mathrm{,}
\end{equation}
where $a = 6.37\times10^6~\mathrm{m}$ is the planetary radius and $H$ is the smaller of the height of the tropopause and scale height averaged between $30^\circ-80^{\circ}$ latitude. The ratio of partial derivatives of potential temperature with respect to latitude and height gives the slope of isentropes from our simulations. We average these partial derivatives from a pressure of $0.5 p_s$ (where $p_s$ is the surface pressure) to the surface and between $30^\circ-80^{\circ}$ latitude. \Eq{eq:criticalityfromgcm} calculates the criticality parameter as an average of local isentropic slopes, which is not equivalent to the ratio of the equator-to-pole temperature contrast and bulk lapse rate. We describe the calculation of the equator-to-pole temperature contrast and bulk lapse rate from our GCM experiments in \Sec{sec:tempcontrasts}.    \\
\indent \Fig{fig:crit} shows a comparison of the baroclinic criticality parameter from our GCM results (solid lines) with the scaling theory from \Eq{eq:xiscaling} (dashed lines) for varying rotation rate and surface pressure. In this comparison, we set $\xi = \xi_\varoplus$ for an Earth-like rotation rate and surface pressure. We find that the scaling of criticality parameter as $\Omega^{2/5}$ from \Eq{eq:xiscaling} is in good agreement with simulations that have rotation rates $\Omega \ge 0.5 \Omega_\varoplus$ (left panel of \Fig{fig:crit}). The under-prediction of the criticality parameter at slower rotation rates is not surprising, as planets with rotation rates of $\Omega \lesssim 1/3 \Omega_\varoplus$ have weak baroclinic effects because the deformation radius is larger than the planetary circumference. \\
\indent We also find good agreement between our theoretical scaling of the criticality parameter with surface pressure and GCM experiments (right hand panel of \Fig{fig:crit}). Note that for the scaling with surface pressure, we include the minimum of the tropopause height and scale height calculated from our GCM experiments in the predicted criticality, because the tropopause height increases with increasing surface pressure. As a result, the decrease of baroclinic criticality with increasing surface pressure is due to the combined effects of increasing surface pressure and increasing tropopause height. 
\subsection{Equator-to-pole temperature contrast and bulk lapse rate}
\label{sec:tempcontrasts}
\indent \Fig{fig:delta} shows a comparison of the equator-to-pole temperature contrast in the GCM experiments (solid lines) with our theory from Equation (\ref{eq:deltahscaling}) (dashed lines) for varying rotation rate and surface pressure. In the GCM results, we average the equator-to-pole temperature contrast from $0.5 p_s$ to the surface. We keep $\Delta_h \theta_\mathrm{eq}$ fixed at $242 \ \mathrm{K}$, which is the radiative equilibrium emission temperature contrast between the equator and the pole for a zero obliquity planet with Earth's incident stellar flux. Our estimate of $\Delta_h \theta_\mathrm{eq}$ assumes an albedo of $0.55$, which is the albedo of our simulation with an Earth-like value of rotation rate and surface pressure. In keeping $\Delta_h \theta_\mathrm{eq}$ fixed, we ignore changes in the albedo due to changes in the sea ice cover. Notice that, with this choice, there is no tunable parameter in our equations for the equator-to-pole temperature contrast and bulk lapse rate. \\
\indent We find that the theoretical scalings for the equator-to-pole temperature contrast broadly match those from our GCM experiments, although not as well as for the criticality parameter. The equator-to-pole temperature contrast increases with increasing rotation rate and decreasing surface pressure in our GCM experiments, due to the increased criticality parameter leading to reduced eddy length scales. Note that the equator-to-pole temperature contrast is greatly reduced for simulations with surface pressures of 2 and 4 bars because they are in an ice-covered state due to the lack of CO$_2$. The change from partial to total ice cover causes a change in $\Delta_h\theta_\mathrm{eq}$, which is not accounted for in our prediction. \\
\indent \Fig{fig:bulk} shows the bulk lapse rate calculated from our suite of GCM experiments, averaged between $30-80^\circ$ latitude, along with predictions from our theoretical scaling (Equation \ref{eq:deltavscaling}). We find that the bulk lapse rate does not show as strong of a dependence on planetary parameters as the equator-to-pole temperature contrast because our simulations are all near the $\xi \approx 1$ regime. Our theoretical prediction for the bulk lapse rate increases with increasing rotation rate throughout the parameter regime studied, with a maximum at $\approx 8 \Omega_\varoplus$. This is because $\xi < 1$ in our simulations for rotation rates $\Omega < 8 \Omega_\varoplus$ (see \Fig{fig:crit}), and when $\xi < 1$ the bulk lapse rate scales with the criticality parameter (see Equation \ref{eq:bulklapse}). Our prediction for the bulk lapse rate is not strongly dependent on pressure, but does have a maximum at a pressure around $0.5 \ \mathrm{bars}$. This maximum at $0.5 \ \mathrm{bars}$ occurs because at this surface pressure $\xi \approx 1$, while at lower pressures the criticality parameter $\xi > 1$ and at higher pressures $\xi < 1$. Our simulations also show a maximum in bulk lapse rate at an intermediate surface pressure, though the maximum occurs at a slightly larger pressure than expected from the theory. 
\section{Discussion \& Conclusions}
\label{sec:disc}


\indent Our predictions for the equator-to-pole temperature contrast and bulk lapse rate can be tested by future observations of terrestrial exoplanet atmospheres with LUVOIR/HabEx/OST. We predict that the equator-to-pole temperature contrast should increase with increasing rotation rate and decreasing surface pressure. Our theoretical predictions can be tested by observationally constraining the equator-to-pole temperature contrast. It may be possible to constrain the equator-to-pole temperature contrast for planets with non-zero obliquity, as both their polar and equatorial regions are visible throughout one planetary rotation \citep{Cowan:2018aa,Olson:2018aa}. Our scalings show that, when varying only rotation rate and considering fixed incident stellar flux, surface pressure, and atmospheric composition, planets with faster rotation rates will have colder poles. As a result, planets in a cold climate regime like those simulated in this work should have wider ice coverage with increasing rotation rate, which could be detectable through albedo variations over planetary orbital phase. \\
\indent We predict that the bulk lapse rate is only weakly dependent on planetary parameters relative to the equator-to-pole temperature contrast. We also found that the bulk lapse rate has a maximum for planets with baroclinic criticality parameters $\xi \approx 1$. For Earth-like planets with rotation rates $\Omega \lesssim 8 \Omega_\varoplus$, we find in our GCM experiments that the bulk lapse rate increases with rotation rate, while for faster rotating planets the bulk lapse rate should decrease with increasing rotation rate. We also find that the bulk lapse rate is very weakly dependent on surface pressure, with less than a factor of two variation when increasing the surface pressure from 0.25 to 4 bars. The bulk lapse rate is related to the vertical potential temperature contrast. Planets with greater vertical potential temperature contrasts have more stably stratified atmospheres with smaller in-situ temperature lapse rates. As a result, a large bulk (potential temperature) lapse rate implies a small in-situ temperature lapse rate. In this way, our predictions for the bulk lapse rate can be tested with inverse (retrieval) methods that constrain the atmospheric lapse rate from an observed spectrum, which have been applied widely to interpret previous exoplanet observations \citep{Madhusudhan:2009,Benneke:2012,Line:2013,Waldmann:2015aa,Feng:2016aa}. Retrieval methods can determine the height of the top of the tropospheric cloud layer \citep{Feng:2018aa}, which occurs approximately at the tropopause. Given the height of the tropopause, we can relate the bulk lapse rate directly to the temperature lapse rate and compare it to our theoretical prediction.  \\
\indent We tested our analytic theory using simulations of terrestrial exoplanets with varying rotation rate and surface pressure. However, this suite of simulations was limited in extent, as we kept the atmospheric composition and host stellar spectrum fixed. We only considered an N$_2$-H$_2$O atmosphere without additional greenhouse gases,  and as a result our simulations are in a relatively cold climate regime with large sea-ice coverage, relatively little water vapor, and weak latent heat transport. If our simulations were instead in a warm climate regime, for example by including an Earth-like complement of greenhouse gases, the equator to pole temperature contrast would be smaller \citep{Kaspi:2014,Jansen:2018aa}. Additionally, we used the Solar incident spectrum for our suite of simulations. Near-IR absorption by water vapor could change the bulk lapse rate from that predicted by our theory for non-tidally locked planets that orbit later spectral type host stars than the Sun \citep{Cronin:2016}. \\
\indent Our theoretical model, while applicable throughout a wide range of parameter space, can only be applied to predict the atmospheric circulation of planets that have baroclinically unstable atmospheres. Baroclinic effects are significantly weaker on slowly rotating planets that have deformation radii which are larger than the planetary circumference. As a result, our theory does not apply to Earth-sized planets with rotation periods $\gtrsim 3~\mathrm{days}$. Note that Earth-like planets could have a wide range of spin rates, with rotation periods ranging from $1-10^4 \ \mathrm{hr}$ due to the stochastic nature of the oligarchic stage of planetary growth \citep{Miguel:2010aa}. Additionally, Earth itself rotated faster in the past, as its rotation is slowed due to tidal interactions with the Moon. The theory presented here cannot be applied to slowly-rotating terrestrial planets orbiting Sun-like stars or tidally locked planets orbiting M dwarf stars. However, \cite{Wordsworth:2014} and \cite{Koll:2016} have developed theories for the temperature structure and wind speeds of tidally locked terrestrial planets. Additionally, theories developed to understand circulation in the tropical regions of Earth \citep{Held:1980,held:2000,Sobel:2001,chemke:2017}
can be used to understand the atmospheric heat transport of slowly rotating planets orbiting Sun-like stars. \\
\indent Our scaling theory could also help with the interpretation of observations of warm Jupiter and warm Neptune exoplanets with the James Webb Space Telescope (JWST). Baroclinic instabilities play a key role in driving the zonal jets in Jupiter's atmosphere \citep{Williams:1978aa,Gierasch:1979aa,Williams:1979aa,Kaspi:2006aa,Lian:2008aa,Young:2019aa}, and likely affect the circulation of fast-rotating gas giant exoplanets. Previous work has studied how the atmospheric circulation of warm Jupiters depends on rotation rate, incident stellar flux, and obliquity \citep{Showman:2014,Rauscher:2017aa}. \cite{Showman:2014} found that the equator-to-pole temperature contrast of warm Jupiters increases with increasing rotation rate, as expected from our scaling theory. Emission spectra with JWST will constrain the temperature-pressure profiles of warm Jupiters and warm Neptunes \citep{Greene:2015}, which may also be affected by baroclinic instabilities. \\
\indent In this work, we extended Earth-based theoretical scalings for baroclinic instability to estimate basic quantities of the circulation of terrestrial exoplanets. Our scalings can be used to predict the baroclinic criticality parameter throughout the baroclinically unstable regime of terrestrial exoplanets. As the criticality parameter is directly linked to the slope of isentropes, we apply these scalings to estimate the equator-to-pole temperature contrast and bulk lapse rate. These scalings predict that the equator-to-pole temperature contrast increases with increasing rotation rate and decreasing surface pressure, while the bulk lapse rate depends relatively weakly on variations in planetary parameters around Earth-like values and has a maximum when the criticality parameter is near one. We find reasonable agreement between the equator-to-pole temperature contrast and bulk lapse rate predicted by our scaling theory and simulated using a detailed GCM. This agreement extends throughout the parameter regime of Earth-sized exoplanets with rotation rates $\gtrsim 1/3 \Omega_\varoplus$, but for more slowly rotating planets baroclinic effects are small and our theory no longer applies. 
\acknowledgements
We thank the referee for helpful comments that improved the manuscript. This work was supported by the NASA Astrobiology Program Grant Number 80NSSC18K0829 and benefited from participation in the NASA Nexus for Exoplanet Systems Science research coordination network. T.D.K. acknowledges funding from the 51 Pegasi b Fellowship in Planetary Astronomy sponsored by the Heising-Simons Foundation. Our work was completed with resources provided by the University of Chicago Research Computing Center. 

\if\bibinc n
\bibliography{References_terrestrial}
\fi

\if\bibinc y

\fi

\end{document}